\documentclass[aip,jcp,reprint,amsmath,amssymb,floatfix,nofootinbib]{revtex4-2}

\usepackage{enumitem}
\usepackage{chemformula}
\usepackage{graphicx}
\usepackage{bm}
\usepackage{dcolumn}
\usepackage{color}
\usepackage{xspace}
\usepackage{hyperref}
\usepackage{booktabs}

\makeatletter
\renewcommand{\@biblabel}[1]{#1.~}

\makeatother

\begin{document}

\title{Quantum Boltzmann Equation Self-Consistent-Field for the Entropic Regularization of Mean-Field Singularities}

\author{Romit Chakraborty}
\email[Author to whom correspondence should be addressed: ]{romit@pointreyessound.com}
\affiliation{Point Reyes Sound, Inc., San Francisco, California 94114, USA}

\date{August 14, 2026}

\begin{abstract}
We present a Quantum Boltzmann Equation self-consistent-field (QBE-SCF) formulation for molecular electronic structure in which the one-electron reduced density matrix is propagated in an atomic orbital basis and relaxed by a Bhatnagar-Gross-Krook collision operator toward a Fermi-Dirac equilibrium defined by the instantaneous Fock matrix. At stationarity, the converged density and Fock matrices satisfy $[\mathbf{F},\mathbf{P}]=0$, the Hartree-Fock condition. While the zero-temperature equilibrium target reduces to the integer Aufbau projector, the damped collision operator ensures the steady-state density matrix is not necessarily idempotent. This kinetic relaxation affords a dual pathway to resolve mean-field singularities. For spatial degeneracies, such as H$_3$ symmetric dissociation, zero-temperature kinetic ergodicity fractionalizes the active space to recover the Generalized Valence Bond (GVB) limit. For the conical intersection in BeH$_2$ and the H$_4$ structural distortion ($D_{2h} \rightarrow D_{4h} \rightarrow D_{2h}$), finite-temperature entropic regularization recovers correlated adiabatic surfaces from a real-valued single-reference density. By maintaining stable numerical convergence across basis-set hierarchies and resolving static correlation without multi-reference wavefunctions, these results establish kinetic relaxation as a synthesis of single-reference electronic structure and quantum statistical mechanics. 
\end{abstract}

\maketitle
\setcounter{footnote}{0}
\renewcommand{\thefootnote}{\fnsymbol{footnote}}

\section{Introduction}
Conventional electronic structure theory treats the ground state as a static solution to a matrix eigenvalue problem, distinct from phase-space formulations that describe quantum mechanics as a dynamical transport process.\cite{Subotnik2026, Wigner1932} The latter perspective proves essential for resolving the derivative discontinuities and artificial symmetry breaking inherent to the mean-field approximation.\cite{SubotnikBeH2} Reframing the electron cloud in phase-space rather than as a fixed wavefunction in real space enables the recovery of correlation and topology lost in standard mean-field methods. While pioneering efforts successfully mapped the Kohn-Sham equations onto lattice grids\cite{Mendoza2013, solorzano2016second} within the Lattice Boltzmann framework,~\cite{Succi2001} these approaches struggle to resolve the sharp nuclear cusps or attain the efficiency required to compete with modern quantum chemistry solvers.\cite{Epifanovsky2021, PySCF2025} We address these limitations by introducing the Quantum Boltzmann Equation--Self-Consistent field formalism (QBE-SCF), merging kinetic rigor with the efficiency of Gaussian basis sets. 

In this work, we use a BGK relaxation scheme in an atomic-orbital (AO) basis to connect phase-space transport to standard self-consistent-field electronic structure. We demonstrate: (i) stable convergence to the exact Restricted Hartree-Fock limit across a basis-set hierarchy; (ii) a kinetic interpretation of $\text{H}_2$ dissociation, where the Coulson-Fischer instability coincides with the spontaneous onset of spatial entropy to recover the Mott limit; (iii) zero-temperature kinetic ergodicity that fractionalizes the active space to bypass the spatial degeneracy in $\text{H}_3$; and (iv) finite-temperature thermodynamic optimization that utilizes configurational entropy to escape the diabatic traps of the $\text{H}_4$ adiabatic loop and the $\text{BeH}_2$ conical intersection. QBE-SCF and all associated algorithmic implementations presented herein were developed at Point Reyes Sound, Inc.

\section{THEORETICAL FRAMEWORK}

\subsection{Phase-Space Dynamics in the AO Basis}
We employ a phase-space formulation for wavefunction mechanics via the Wigner quasi-probability distribution, $W(\mathbf{r},\mathbf{p},t)$.~\cite{Wigner1932, hillery1984wigner} To render this continuous function computationally tractable for molecular systems, we project $W(\mathbf{r},\mathbf{p},t)$ onto a finite basis of atomic orbitals, $\{\chi_\mu\}$:~\cite{Subotnik2026}
\begin{equation}
W(\mathbf{r},\mathbf{p},t) = \sum_{\mu\nu} P_{\mu\nu}(t) \mathcal{W}_{\mu\nu}(\mathbf{r},\mathbf{p}) \label{eq:wigner_ansatz}
\end{equation}
where $P_{\mu\nu}(t)$ is the time-dependent one-particle density matrix and $\mathcal{W}_{\mu\nu}$ is the Wigner transform of the orbital dyad $|\chi_\mu\rangle\langle\chi_\nu|$. The underlying Gaussian basis functions are locked to stationary nuclear centers under the Born-Oppenheimer approximation. 

The evolution of this projected density is governed by the continuous Quantum Boltzmann Equation:
\begin{equation}
\frac{\partial W}{\partial t} + \frac{\mathbf{p}}{m} \cdot \nabla_\mathbf{r} W + \mathcal{F} \cdot \nabla_\mathbf{p} W = \mathcal{Q}[W] \label{eq:qbe_continuous}
\end{equation}
where the streaming of probability density is balanced by a collision term, $\mathcal{Q}[W]$, driving the fluid towards thermodynamic equilibrium.~\cite{Subotnik2026, Mendoza2013} 

To translate this phase-space advection back into our discrete atomic orbital basis, we invoke the Wigner-Weyl transform.~\cite{weyl1927quantenmechanik, zachos2005quantum} Under the mean-field approximation the many-body Hamiltonian may be replaced by the single-particle Fock operator $^{1}\mathbf{F} = \mathbf{T} + \mathbf{V}_{\mathrm{eff}}$, and the exact density ${}^{N}\mathbf{D}$ with our projected one-particle density matrix ${}^{1}\mathbf{P}$. The continuous Quantum Boltzmann Equation then maps to the quantum Liouville equation in matrix form:
\begin{equation}
\frac{\partial \mathbf{P}}{\partial t} - i[\mathbf{F}, \mathbf{P}] = \mathcal{Q}[\mathbf{P}] \label{eq:liouville}
\end{equation}

\subsection{BGK Relaxation}
To model the collision operator $\mathcal{Q}[\mathbf{P}]$ in Equation \ref{eq:liouville}, we employ the continuous Bhatnagar-Gross-Krook (BGK) relaxation time approximation.~\cite{Bhatnagar1954} This approximates collisions as a statistical friction that drives the non-equilibrium density matrix toward a local target equilibrium, $\mathbf{P}^{\mathrm{eq}}$, over a relaxation time $\tau$.

Since our objective is to isolate the stationary ground state, we drop the commutator during the micro-step update since integrating the unitary rotation term, $-i[\mathbf{F}, \mathbf{P}]$. This isolates the solver as a stable optimizer in imaginary time, rather than a real-time dynamics simulator and is justified since the target equilibrium $\mathbf{P}^{\mathrm{eq}}$ is constructed from the eigenvectors of the instantaneous Fock matrix and satisfies $[\mathbf{F}, \mathbf{P}^{\mathrm{eq}}] = \mathbf{0}$. 

By discretizing the temporal relaxation and defining the dimensionless collision frequency as $\omega = \frac{\Delta t}{\tau}$, we arrive at the discrete matrix propagation scheme derived explicitly in the Supplementary Information:
\begin{equation}
\mathbf{P}(t + \Delta t) = (1 - \omega)\mathbf{P}(t) + \omega \mathbf{P}^{\mathrm{eq}}[\mathbf{F}(t)] \label{eq:bgk_discrete}
\end{equation}

Equation \ref{eq:bgk_discrete} casts the self-consistent-field update as a local, Markovian relaxation process. In the undamped limit of $\omega = 1.0$, this reduces to the memoryless Roothaan repeated-diagonalization step, which frequently diverges due to non-linear charge sloshing. Enforcing $\omega < 1.0$ retains the preceding density matrix, and damps these oscillations. Consequently, QBE-SCF stabilizes the mean field through instantaneous kinetic friction, bypassing the need for non-Markovian convergence accelerators like DIIS that require storing, evaluating, and extrapolating from extensive arrays of prior state vectors.

At a converged fixed point, this damped iteration satisfies the Hartree-Fock self-consistency condition. Stationarity ($\mathbf{P}_{t+\Delta t} = \mathbf{P}_{t}$) implies the condition $\mathbf{P}_{\mathrm{stat}} = \mathbf{P}^{\mathrm{eq}}[\mathbf{F}(\mathbf{P}_{\mathrm{stat}})]$. Since the equilibrium distribution $\mathbf{P}^{\mathrm{eq}}$ is constructed via Fermi-Dirac statistics, it shares the eigenvectors of the Fock matrix $\mathbf{F}$. Consequently, the stationary density satisfies:
\begin{equation}
[\mathbf{F}_{\mathrm{stat}}, \mathbf{P}_{\mathrm{stat}}] = \mathbf{0}
\end{equation}
This commutation relation represents the exact Hartree-Fock stationarity condition. In the zero-temperature limit ($T_{\mathrm{elec}} \rightarrow 0$), the target equilibrium $\mathbf{P}^{\mathrm{eq}}$ reduces to the integer Aufbau projector. Standard single-reference solvers enforce idempotency ($\mathbf{P}^2 = \mathbf{P}$) at each step, thereby restricting the system to zero-entropy diabatic states that force non-differentiable cusps at degeneracies. In contrast, QBE-SCF propagates the density matrix via a convex combination ($\omega < 1.0$). Because any convex combination of distinct idempotent matrices is non-idempotent, the converged steady state $\mathbf{P}_{\mathrm{stat}}$ is not necessarily restricted to idempotency. As we will demonstrate, this allows the zero-temperature density matrix to spontaneously fractionalize across degenerate manifolds, generating spatial entropy to bypass the topological traps of conventional mean-field theory.

\subsection{Regularization of Mean-Field Singularities}
For molecular configurations exhibiting strong static correlation, idempotency-constrained solvers frequently collapse into symmetry-broken solutions, yielding unphysical derivative discontinuities (cusps) on the potential energy surface. QBE-SCF provides two distinct physical mechanisms to regularize these mean-field singularities, depending on the underlying molecular topology.

For true spatial degeneracies dictated by point-group symmetry, regularization emerges natively at the zero-temperature Aufbau limit ($T_{\mathrm{elec}} = 0$). While repeated-diagonalization schemes trap the density matrix in a localized, broken-symmetry state, the Markovian collision operator ($\omega < 1.0$) continuously dampens the update, incorporating a fraction of the asymmetric projection into the global state. This localized charge injection biases the spatial potential landscape, driving a sequential unitary rotation of the frontier eigenvectors. Rather than sampling a thermal ensemble, the temporal accumulation of these discrete, integer-occupied updates generates a kinetic ergodicity over the degenerate subspace. This fractionalizes the steady-state density matrix, allowing the solver to bypass the single-determinant derivative discontinuity and track the adiabatic limit without invoking thermal energy.

Conversely, when the mean-field cusp arises from an artificial diabatic trap rather than a true spatial degeneracy such as an avoided crossing of states with identical symmetry—this zero-temperature kinetic ergodicity is insufficient to bridge the gap. We resolve this barrier through thermodynamic regularization by introducing a finite electronic temperature ($T_{\mathrm{elec}}=\beta^{-1}>0$) and minimizing the macroscopic Helmholtz free energy ($F=E-T_{\mathrm{elec}}S_{\mathrm{vN}}$) rather than the internal energy alone.

In this regime, the microscopic von Neumann entropy of the system is evaluated across the independent spin channels as $S_{\mathrm{vN}} = -\sum_{\sigma \in \{\alpha, \beta\}} \mathrm{Tr} \left[ \mathbf{P}^\sigma \ln \mathbf{P}^\sigma + (\mathbf{I} - \mathbf{P}^\sigma) \ln (\mathbf{I} - \mathbf{P}^\sigma) \right]$. For entropy plots expressed in bits, the thermodynamic von Neumann entropy driving the QBE-SCF minimization is defined as $S_{\mathrm{vN}}$(bits)$=\frac{S_{\mathrm{vN}}}{\ln 2}$. Because standard Fermi-Dirac statistics assumes independent quasi-particles, this formulation includes ionic charge fluctuations (e.g., $\text{H}^+\text{H}^-$), causing the entropy of symmetrically dissociated $\text{H}_2$ to evaluate to 4.0 bits. To quantify spin entanglement and recover the classical Mott insulator limit (2.0 bits for two isolated hydrogen atoms), we define a configurational spin entropy based on the fractional spin-orbital occupation, $n_i = p_i / 2$ where $p_i$ denotes the eigenvalue of the total spatial density matrix: 
\begin{equation}
S_{\mathrm{config}}(\mathrm{bits}) = -\sum_{i} \left[ n_i \log_2(n_i) + (1-n_i) \log_2(1-n_i) \right]
\end{equation}
At the equilibrium geometry, the spatial density matrix is idempotent ($n_i \in \{0, 1\}$), causing the entropy to vanish and correctly recover the closed-shell limit. As the bond dissociates and fractional occupations emerge, this equation filters out the unphysical empty and doubly occupied ionic charge fluctuations suppressed by massive on-site Coulomb repulsion.\cite{hubbard1963electron} This allows the single-particle density to correctly resolve the strongly correlated Mott insulator limit across the entire reaction coordinate.~\cite{hachmann2006multireference, tsuchimochi2009strong}

Eigenfunctions of the instantaneous Fock matrix can be treated as the quasiparticles of a local Fermi liquid, dressed by the collective Coulomb and exchange interactions of the mean field.~\cite{Pines1966} As in standard neutral Fermi liquid transport theory, the macroscopic evolution of the fluid across the singularity is stabilized by the fractional Fermi-Dirac occupation of these specific quasiparticle states. By utilizing entropy as a physical degree of freedom, this thermal smearing transforms non-differentiable energy cusps into smooth thermodynamic crossovers, recovering the correlated multi-reference topology via Fermi-Dirac statistics. Consequently, these quasiparticles possess a population governed by thermodynamics, while their effective potentials are dictated by Coulomb repulsion and exchange interactions.

\section{Computational Methodology}

All simulations were performed within the Born-Oppenheimer approximation using a non-relativistic Hamiltonian, neglecting spin-orbit coupling. The numerical framework was executed using a custom solver developed for this work that integrates the Quantum Boltzmann Equation via the BGK relaxation scheme. The target equilibrium $\mathbf{P}^{\mathrm{eq}}$ is updated self-consistently from the instantaneous Fock matrix, $\mathbf{F}(t) = \mathbf{H}_{core} + \mathbf{J}[\mathbf{P}] - \mathbf{K}[\mathbf{P}]$. All underlying atomic orbital integrals and reference electronic structure solutions were generated using the PySCF package.\cite{PySCF2025}
\begin{figure*}[htbp]
\centering
\includegraphics[width=\textwidth]{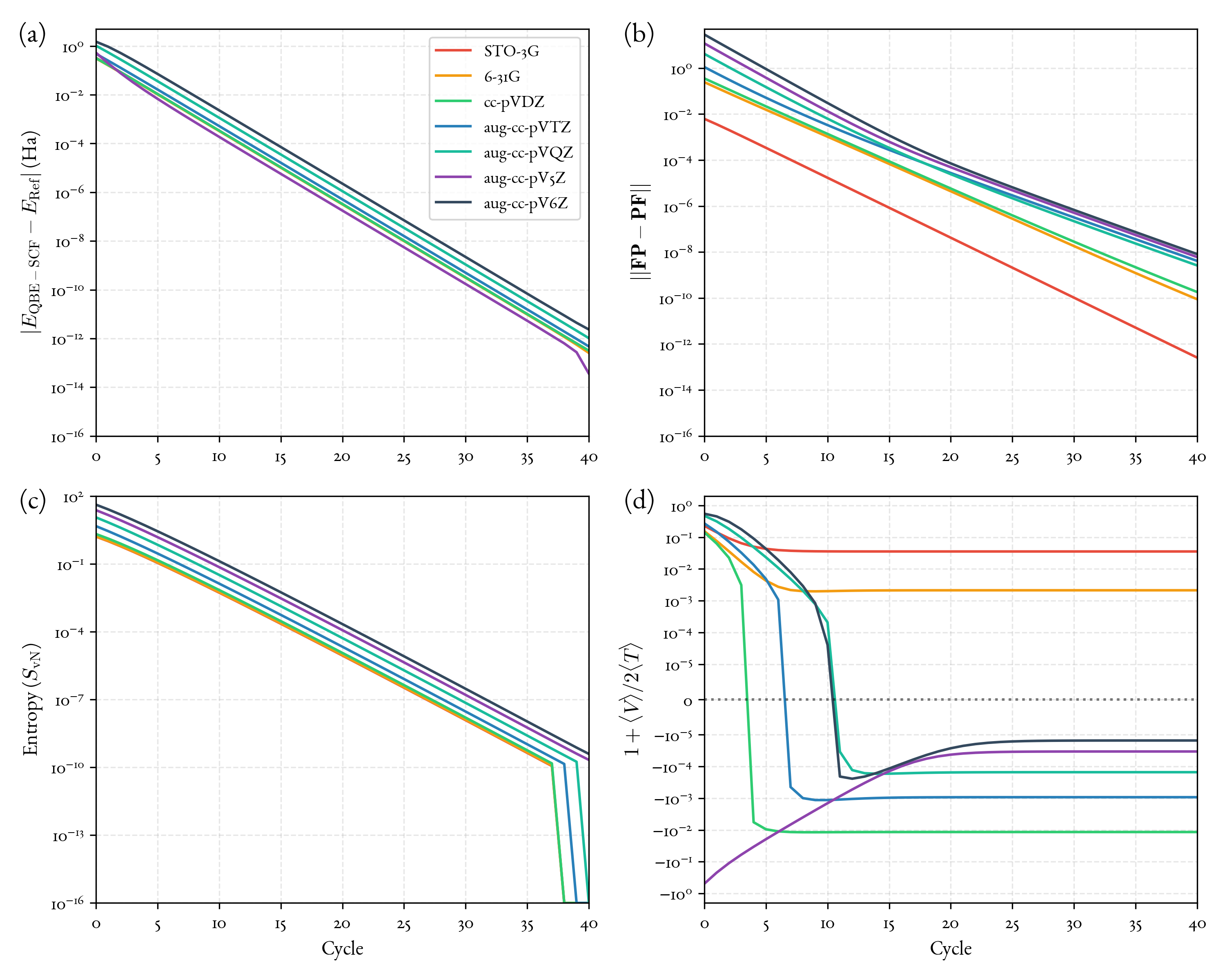}
\caption{\textbf{$\text{H}_2$ at Equilibrium}. QBE-SCF relaxes a perturbed SAD guess across a hierarchy of basis sets. We see   the monotonic decay of energy error w.r.t Restricted Hartree-Fock in \textbf{(a)}, the commutator norm $\| \mathbf{F}\mathbf{P} - \mathbf{P}\mathbf{F} \|$ in \textbf{(b)}. The density converges to a pure state in \textbf{(c)} and \textbf{(d)} confirms restoration of the virial theorem at equilibrium.
}
\label{fig:convergence_dynamics}
\end{figure*}

\subsection{Simulation Parameters}

\begin{description}[wide=0pt]
    \item [Bases]We utilized a hierarchy of basis sets starting from STO-3G to Dunning's correlation-consistent basis sets, ranging from cc-pVDZ to the near-complete limit (aug-cc-pV6Z) to distinguish numerical convergence from physical accuracy. 
    
    \item [Relaxation Time]The collision frequency was set to $\omega = 0.5$ ($\tau = 2.0 \Delta t$), and ensuring monotonic convergence without reliance on non-Markovian extrapolation schemes (e.g., DIIS).
    
    \item [Initial Guess]Simulations were initialized from a perturbed Superposition of Atomic Densities (SAD) with Gaussian noise $\xi \sim \mathcal{N}(0, 0.1)$ to test the solver's robustness unless stated otherwise. 
    
    \item [Thermodynamics] Ground-state benchmarks enforced the $\beta \to \infty$ limit via the Aufbau principle. For thermodynamic regularization studies, a finite electronic temperature (in Ha) was introduced. The chemical potential $\mu$ was determined at each self-consistent-field iteration by finding the root of the particle conservation constraint, $f(\mu) = \sum_{i} n_i(\mu) - N_{\mathrm{elec}} = 0$. This was executed using the \texttt{brentq} root-finding algorithm (as implemented in SciPy) bracketed within the valence eigenvalue spectrum, ensuring convergence to machine precision ($10^{-12}$). Figure~\ref{fig:h2_free} illustrates $\text{H}_{2}$ free energy change as optimized by QBE as a function of bond length.
    
    \item [Convergence]Stationarity was defined by the decay of the commutator norm $||\mathbf{F}\mathbf{P} - \mathbf{P}\mathbf{F}|| < 10^{-7}$. Physical consistency was further verified via the Virial Theorem, wherein the total electronic kinetic energy is evaluated directly in the static atomic orbital basis ($\langle T \rangle = \mathrm{Tr}(\mathbf{P}\mathbf{T})$) to continuously monitor the spatial decompression of the fluid, requiring the metric $1 + \langle V \rangle / 2\langle T \rangle \to 0$ for electronic structure computations at the optimal geometry at the basis set limit.
\end{description}

\section{Results and Discussion}
\subsection{Convergence at Equilibrium in \texorpdfstring{H$_2$}{H2}}

Quantum Boltzmann Equation Self-Consistent-Field (QBE-SCF) was benchmarked against a hierarchy of Dunning's correlation-consistent basis sets, ranging from cc-pVDZ with $L_{max}=p$ to aug-cc-pV6Z with $L_{max}=h$. Figure \ref{fig:convergence_dynamics} tracks the trajectory of the electronic fluid as it relaxes from a perturbed initial state to the variational minimum. The fixed nuclear geometry used for running the QBE-SCF trajectories was optimized at the aug-cc-pV5Z basis with a tight step size of $10^{-6}$ Bohr yielding an internuclear distance of $1.386276$ Bohr.  

We observe stable, monotonic convergence across all metrics. As detailed in Table \ref{tab:h2_convergence}, the energy error relative to exact Restricted Hartree-Fock ($E_{\mathrm{Ref}}$) decays monotonically, reaching microhartree level numerical agreement within 40 iterations (Figure \ref{fig:convergence_dynamics}a) and achieving identical numerical agreement down to the eighth decimal place across the entire basis-set hierarchy. This convergence rate is highly robust against basis set expansion. Scaling the Hilbert space dimensionality from 2 (STO-3G) to 254 (aug-cc-pV6Z) barely impacts the relaxation time, with the solver requiring only between 19 and 35 steps to achieve a commutator norm $||\mathbf{F}\mathbf{P}-\mathbf{P}\mathbf{F}|| < 10^{-7}$ (Figure \ref{fig:convergence_dynamics}b). This demonstrates that instantaneous kinetic friction natively stabilizes the mean field across diverse, high-dimensional landscapes without requiring non-Markovian extrapolation schemes such as DIIS.

\begin{table*}[htbp]
\centering
\caption{\textbf{QBE-SCF convergence for H$_2$ at equilibrium.} Evaluated at collision frequency $\omega=0.5$, the solver maintains flat Cycle counts and exact virial ratios despite massive expansions in the Hilbert space dimension ($N_{\mathrm{orb}}$).}
\label{tab:h2_convergence}
\begin{tabular*}{\textwidth}{@{\extracolsep{\fill}} l c c c c c c c @{}}
\hline\hline
Basis Set & $N_{\mathrm{orb}}$ & $E_{\mathrm{Ref}}\ (E_h)$ & $E_{\mathrm{QBE}}\ (E_h)$ & $||\mathbf{F}\mathbf{P} - \mathbf{P}\mathbf{F}||$ & $-\langle V\rangle/2\langle T\rangle$ & $\mathrm{Tr}(\mathbf{P}\mathbf{S})$ & $\mathrm{Cycles}$ \\
\hline
STO-3G ($s$) & 2 & $-1.117058767$ & $-1.117058109$ & $7.6\times 10^{-8}$ & $0.964328$ & $1.999999994$ & 19 \\
6-31G ($s$) & 4 & $-1.126818241$ & $-1.126818238$ & $9.3\times 10^{-8}$ & $0.997863$ & $2.000000000$ & 27 \\
cc-pVDZ ($p$) & 10 & $-1.128593916$ & $-1.128593915$ & $8.0\times 10^{-8}$ & $1.011600$ & $2.000000000$ & 28 \\
aug-cc-pVTZ ($d$) & 46 & $-1.133055316$ & $-1.133055316$ & $8.6\times 10^{-8}$ & $1.000927$ & $2.000000000$ & 33 \\
aug-cc-pVQZ ($f$) & 92 & $-1.133508496$ & $-1.133508495$ & $8.9\times 10^{-8}$ & $1.000151$ & $2.000000000$ & 32 \\
aug-cc-pV5Z ($g$) & 160 & $-1.133647810$ & $-1.133647810$ & $8.5\times 10^{-8}$ & $1.000034$ & $2.000000000$ & 34 \\
aug-cc-pV6Z ($h$) & 254 & $-1.133663880$ & $-1.133663880$ & $7.3\times 10^{-8}$ & $1.000015$ & $2.000000000$ & 35 \\
\hline\hline
\end{tabular*}
\end{table*}

This energy minimization is accompanied by the rapid purification of the density matrix (Figure \ref{fig:convergence_dynamics}c), confirming that the solver filters out thermal noise to recover the idempotent ground state at equilibrium. Figure \ref{fig:convergence_dynamics}d reveals the physical relaxation mechanism of the electron density. For basis sets containing diffuse functions (cc-pVDZ and larger), the virial error initially plunges below zero. This corresponds to a decompression phase: the initial guess, constructed from isolated atomic orbitals, is spatially too confined  and has high kinetic energy. The solver resolves this by expanding the electron density into the diffuse subspace, lowering the kinetic pressure until the virial theorem condition ($\langle V \rangle = -2\langle T \rangle$) is restored. 

This non-monotonic relaxation of the wavefunction highlights the solver's ability to navigate the complex optimization landscape of high-dimensional Hilbert spaces. The density for STO-3G, and 6-31G, for which $L_{max} = s$ start from a higher relative $\langle K \rangle$ that is maintained throughout their trajectories. For aug-cc-PV5Z, for which the nuclear geometry was optimized, we see that the perturbed SAD guess has a relatively higher potential energy that gets redistributed into the kinetic term as the density relaxes toward idempotency at equilibrium. Convergence of the virial ratio is limited in this instance by the micro-Bohr resolution of the underlying nuclear geometry optimization. As verified in Table \ref{tab:h2_convergence}, the most complete basis utilized (aug-cc-pV6Z) yields a virial ratio of 1.000015 at convergence. In the following figures $E_{QBE}$ and $F_{\mathrm{QBE}}(T)$ refer to the QBE-SCF optimized internal energies and free energies.

\subsection{\texorpdfstring{H$_2$}{H2} Dissociation}

The dissociation of H$_2$ (\texttt{cc-pVTZ}) evaluates the solver's navigation of static correlation. Unrestricted Hartree-Fock (UHF) recovers the correct asymptote by breaking spatial symmetry at the Coulson-Fischer point ($R \approx 1.21\ \mathrm{\AA}$); see Section SIII of the Supplementary Information for a high-resolution grid validation of this critical slowing down).~\cite{coulson1949notes}
\begin{figure}[b!]
    \centering
    \includegraphics[width=\columnwidth]{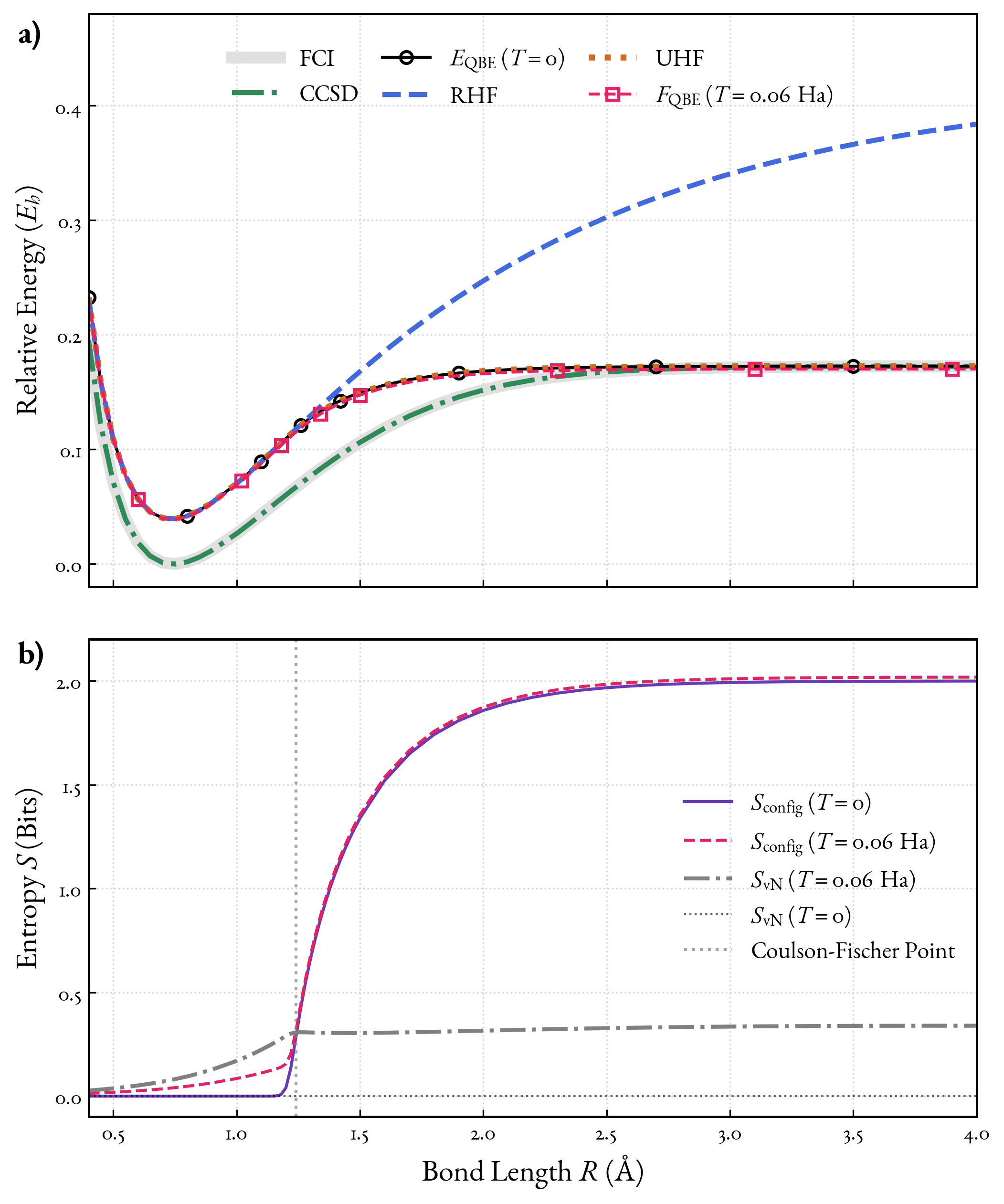}
    \caption{\textbf{H$_2$ Dissociation in \texttt{cc-pVTZ}.} \textbf{(a)} $E_{\mathrm{QBE}}(T=0)$ recovers the UHF asymptote. \textbf{(b)} Configurational entropy ($S_{\mathrm{config}}$) shows transition from a closed-shell singlet ($0.0$ bits) to the perfect-pairing diradical ($\frac{1}{\sqrt{2}}(|\uparrow\downarrow\rangle - |\downarrow\uparrow\rangle)$). Von Neumann entropy ($S_{\mathrm{vN}}$) represents thermal disorder.}
    \label{fig:h2_dissoc}
    \label{fig:h2_free}
\end{figure}
QBE-SCF was initialized independently of converged UHF densities. Symmetry was broken by applying a trace-conserving perturbation, $\mathbf{\Delta}$, across the spin channels of a Superposition of Atomic Densities (SAD) guess ($\mathbf{P}_{\mathrm{SAD}}$) such that $\mathbf{P}^\alpha = \frac{1}{2}\mathbf{P}_{\mathrm{SAD}} + \mathbf{\Delta}$ and $\mathbf{P}^\beta = \frac{1}{2}\mathbf{P}_{\mathrm{SAD}} - \mathbf{\Delta}$.
This initial imbalance triggers Coulomb repulsion, driving relaxation into the broken-symmetry minimum. 

At $T = 0$, the QBE internal energy ($E_{\mathrm{QBE}}$) recovers the UHF minimum. Applying an electronic temperature ($T = 0.06\ \mathrm{Ha}$) yields a free energy ($F_{\mathrm{QBE}}$) that transitions continuously across the Coulson-Fischer point (Figure \ref{fig:h2_dissoc}).

This decouples static correlation from thermal fluctuations. Configurational entropy ($S_{\mathrm{config}}$) captures macroscopic spatial topology: prior to dissociation ($R < 1.21\ \mathrm{\AA}$), $S_{\mathrm{config}}=0.0$ bits and post-transition this saturates at $2.0$ bits for the localized diradical limit~\cite{roos1980complete}. Conversely, the von Neumann entropy ($S_{\mathrm{vN}}$) isolates microscopic thermal disorder via fractional natural orbital occupancies, remaining independent of bond dissociation.

\subsection{\texorpdfstring{$\text{H}_3$}{H3} Dissociation in \texorpdfstring{D$_{3h}$}{D3h}}

During the symmetric dissociation of $\text{H}_3$ parameterized by equilateral side length $R$, the $a_1^\prime$ and $e^\prime$ orbitals merge into a degenerate manifold near $R \approx 1.52\ \mathrm{\AA}$. Standard single-determinant solvers undergo a Coulson-Fischer instability to minimize Coulomb repulsion, breaking the $D_{3h}$ point-group symmetry to enforce idempotency ($\mathbf{P}^2 = \mathbf{P}$). QBE-SCF bypasses this symmetry breaking at $T=0$ through the damped Markovian relaxation of the density matrix: $\mathbf{P}^{(t+1)} = (1-\omega)\mathbf{P}^{(t)} + \omega \mathbf{P}_{\mathrm{eq}}^{(t)}$. To initiate this spatial polarization, the initial state is seeded with a heavily weighted restricted mean-field density perturbed by an infinitesimal spin-unrestricted component.

\begin{figure}[htbp]
    \centering
    \includegraphics[width=\columnwidth]{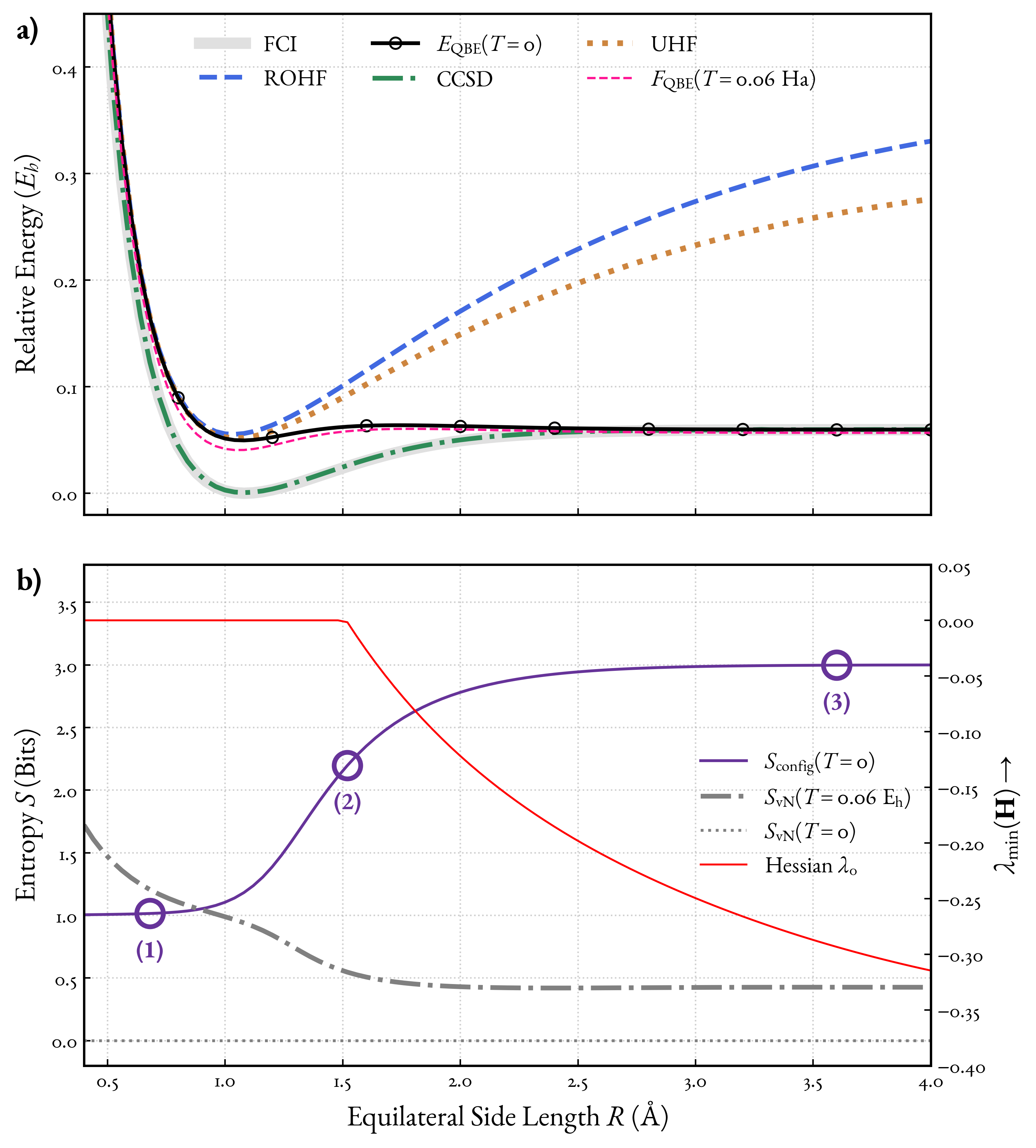}
    \caption{\textbf{$\mathrm{H}_3$ D$_{3h}$ dissociation in \texttt{cc-pVDZ}}. \textbf{(a)} $E_{QBE}(T)=0$ converges to FCI in the dissociative limit as $S_{\mathrm{vN}}$ stays pinned to $0.0$ in \textbf{(b)}. $S_{\mathrm{config}}(T)=0$ maintains a $1.0$-bit baseline for the unpaired radical (1) before instability in the idempotent density ($\lambda_{\mathrm{min}}(\mathbf{H}) < 0$) (2) triggers fractionalization that saturates at $3.0$ bits to recover the Mott insulator limit (3). Natural orbitals at (1), (2), and (3) are shown in Fig.~\ref{fig:h3_orbital_grid}.}
    \label{fig:h3_master}
\end{figure}

By incorporating only a fraction ($\omega$) of the integer Aufbau projection ($\mathbf{P}_{\mathrm{eq}}^{(t)}$), the resulting asymmetric Coulomb penalty directionally biases the potential landscape. This repels the frontier eigenvectors, driving a chronological unitary rotation across the degenerate manifold in imaginary time. The solver integrates this trajectory as an exponential moving average:
\begin{equation}
    \mathbf{P}^{(\infty)} = \lim_{t \to \infty} \omega \sum_{\tau=0}^{t} (1-\omega)^{t-\tau} \mathbf{P}_{\mathrm{eq}}^{(\tau)}
    \label{eq:qbe_steady_state}
\end{equation}
Because any convex combination of distinct idempotent matrices is non-idempotent, this zero-temperature kinetic ergodicity ensures an exact, fractional steady state ($\mathbf{P}^{(\infty)2} \neq \mathbf{P}^{(\infty)}$). This allows QBE-SCF to track the  FCI dissociation limit without thermal smearing (Figure \ref{fig:h3_master}).

\begin{figure*}[htbp]
    \centering
    \setlength{\fboxsep}{0pt}    % Removes padding inside the frame
    \setlength{\fboxrule}{0.5pt} % Sets the frame border thickness
    \setlength{\tabcolsep}{8pt}  % Widens the horizontal whitespace between columns
    \begin{tabular}{c c c c}
        % --- ROW 1: 0.68 A ---
        \shortstack{\textbf{(1)} \\ $R=0.68\ \text{\AA}$\hspace{10pt}} &
        \raisebox{-0.5\height}{\fbox{\includegraphics[width=0.18\textwidth]{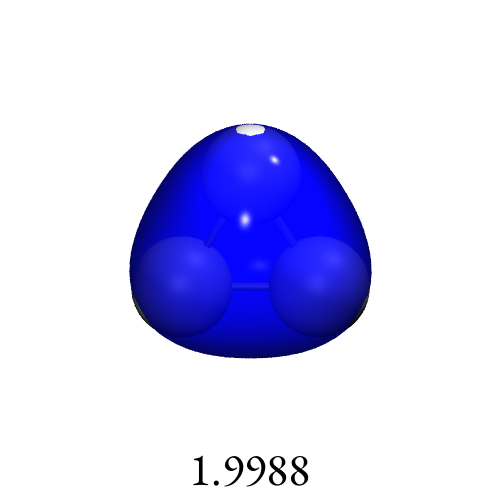}}} &
        \raisebox{-0.5\height}{\fbox{\includegraphics[width=0.18\textwidth]{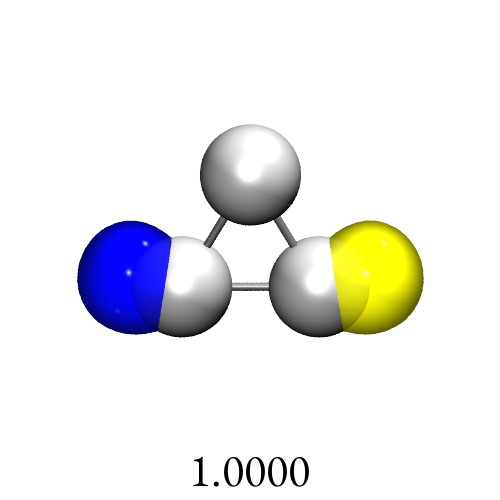}}} &
        \raisebox{-0.5\height}{\fbox{\includegraphics[width=0.18\textwidth]{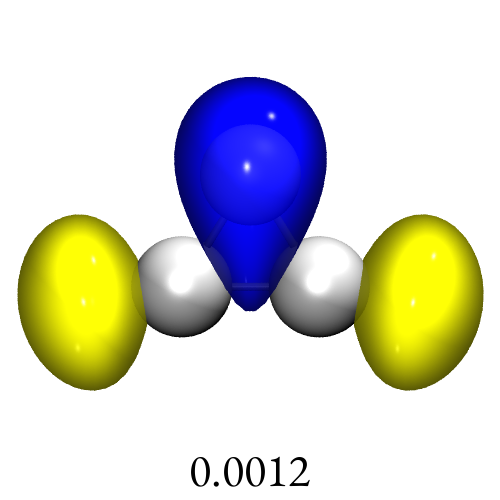}}} \\[25pt]
        
        % --- ROW 2: 1.52 A (The Critical Point) ---
        \shortstack{\textbf{(2)} \\ $R=1.52\ \text{\AA}$\hspace{10pt}} &
        \raisebox{-0.5\height}{\fbox{\includegraphics[width=0.18\textwidth]{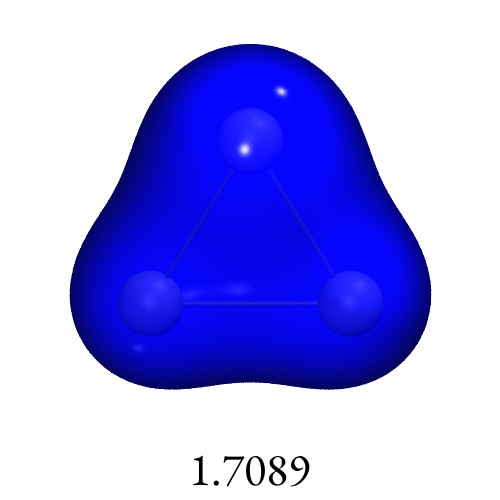}}} &
        \raisebox{-0.5\height}{\fbox{\includegraphics[width=0.18\textwidth]{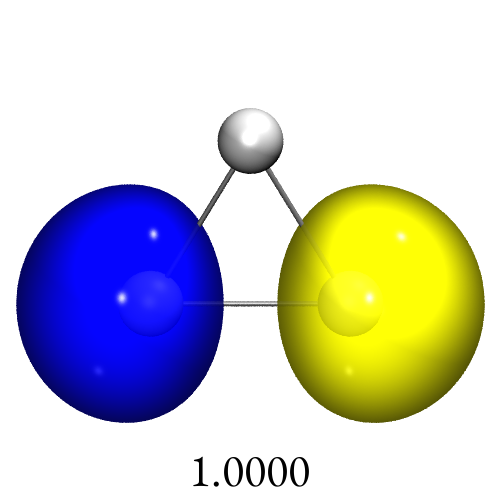}}} &
        \raisebox{-0.5\height}{\fbox{\includegraphics[width=0.18\textwidth]{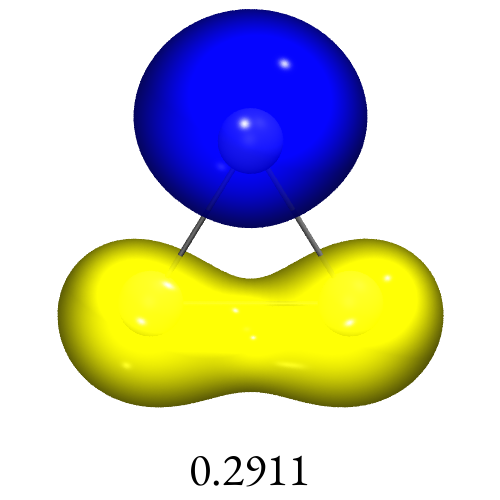}}} \\[25pt]
        
        % --- ROW 3: 3.60 A ---
        \shortstack{\textbf{(3)} \\ $R=3.60\ \text{\AA}$\hspace{10pt}} &
        \raisebox{-0.5\height}{\fbox{\includegraphics[width=0.18\textwidth]{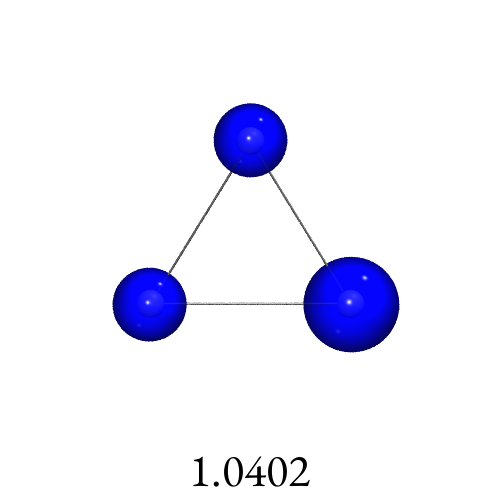}}} &
        \raisebox{-0.5\height}{\fbox{\includegraphics[width=0.18\textwidth]{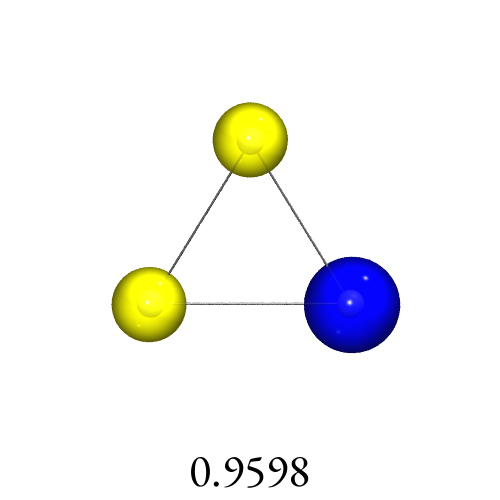}}} &
        \raisebox{-0.5\height}{\fbox{\includegraphics[width=0.18\textwidth]{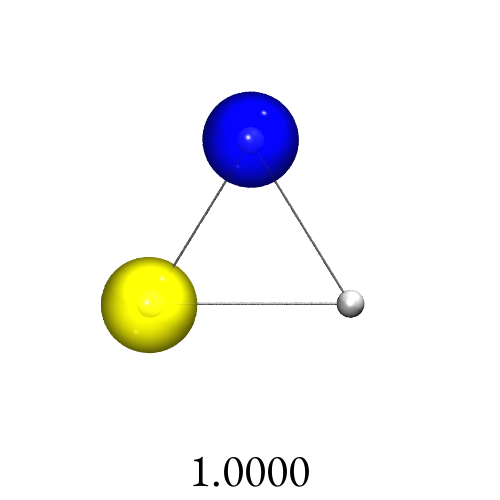}}} \\[15pt]
        
        % Column Labels
        & 
        \makebox[0.18\textwidth]{\textbf{(a)} $p_1$ ($a_1^\prime$)} & 
        \makebox[0.18\textwidth]{\textbf{(b)} $p_2$ ($e^\prime$)} & 
        \makebox[0.18\textwidth]{\textbf{(c)} $p_3$ ($e^\prime$)} \\
    \end{tabular}
    \caption{\textbf{Evolution of the QBE-SCF spatial natural orbitals at $T=0$ during $\text{H}_3$ dissociation.} In the tightly bound molecular regime \textbf{(1)}, the natural orbitals reflect a delocalized closed-shell configuration with a dominant core pair. As internuclear separation increases in \textbf{(2)}, the primary bonding orbital polarizes as macroscopic occupation shifts into the antibonding state, breaking the $D_{3h}$ symmetry. Upon reaching the asymptotic Mott limit in \textbf{(3)}, the spatial natural orbitals localize into non-interacting atomic spheres with exact integer occupations.}
    \label{fig:h3_orbital_grid}
\end{figure*}

The configurational spin entropy ($S_{\mathrm{config}}$) and spatial Natural Orbitals (Figure \ref{fig:h3_orbital_grid}) quantify the underlying spatial mechanics. In the single-reference molecular limit ($R = 0.68\ \text{\AA}$), the unentangled spectator radical ($n_2=1.0$) pins the entropy baseline at exactly $1.0$ bit, while the core pair resides in the fully symmetric $a_1^\prime$ orbital. 

As the Coulson-Fischer instability ($R=1.52\ \text{\AA}$) is triggered when the lowest orbital rotation Hessian eigenvalue ($\lambda_{\mathrm{min}}(\mathbf{H})$) crosses zero, the kinetic ergodicity described in Eq.~\eqref{eq:qbe_steady_state} unlocks the spatial phase space. The solver selectively polarizes the bonding density into the antibonding state, executing a Generalized Valence Bond (GVB) perfect-pairing transition within the single-particle framework.~\cite{goddard1973generalized, bobrowicz1977gvb}

In the strongly correlated dissociation regime ($R = 3.60\ \text{\AA}$), spatial fractionalization saturates as the macroscopic fluid disentangles into three independent atomic $1s$ spheres. This generates exactly $3.0$ bits, perfectly recovering the finite molecular Mott insulator limit.~\cite{hachmann2006multireference, tsuchimochi2009strong}

\subsection{\texorpdfstring{$\text{H}_4$}{H4} Conical Intersection in \texorpdfstring{D$_{4h}$}{D4h}}

We evaluate the degeneracy of the $\text{H}_4$ molecular square ($D_{4h}$) as an intersection via two distortion coordinates: a linear rectangular stretch and a continuous angular loop, both of which are characterized by $D_{2h} \rightarrow D_{4h} \rightarrow D_{2h}$ transitions at their principal axes where the $D_{2h}$ geometries correspond to rectangular or rhombic distortions. The mean-field derivative discontinuity at the $D_{4h}$ geometry arises from the integer Aufbau constraint, which forces a permutation of spatial orbital occupations as the structural symmetry is lowered. Calculations employ a 6-31G basis. The QBE-SCF solver is initialized using the idempotent restricted mean-field (RHF) density matrix, evaluated independently at each coordinate without spatial symmetry-breaking perturbations.
\begin{figure*}[htbp]
    \begin{minipage}[t]{0.48\textwidth}
        \centering
        \includegraphics[width=\linewidth]{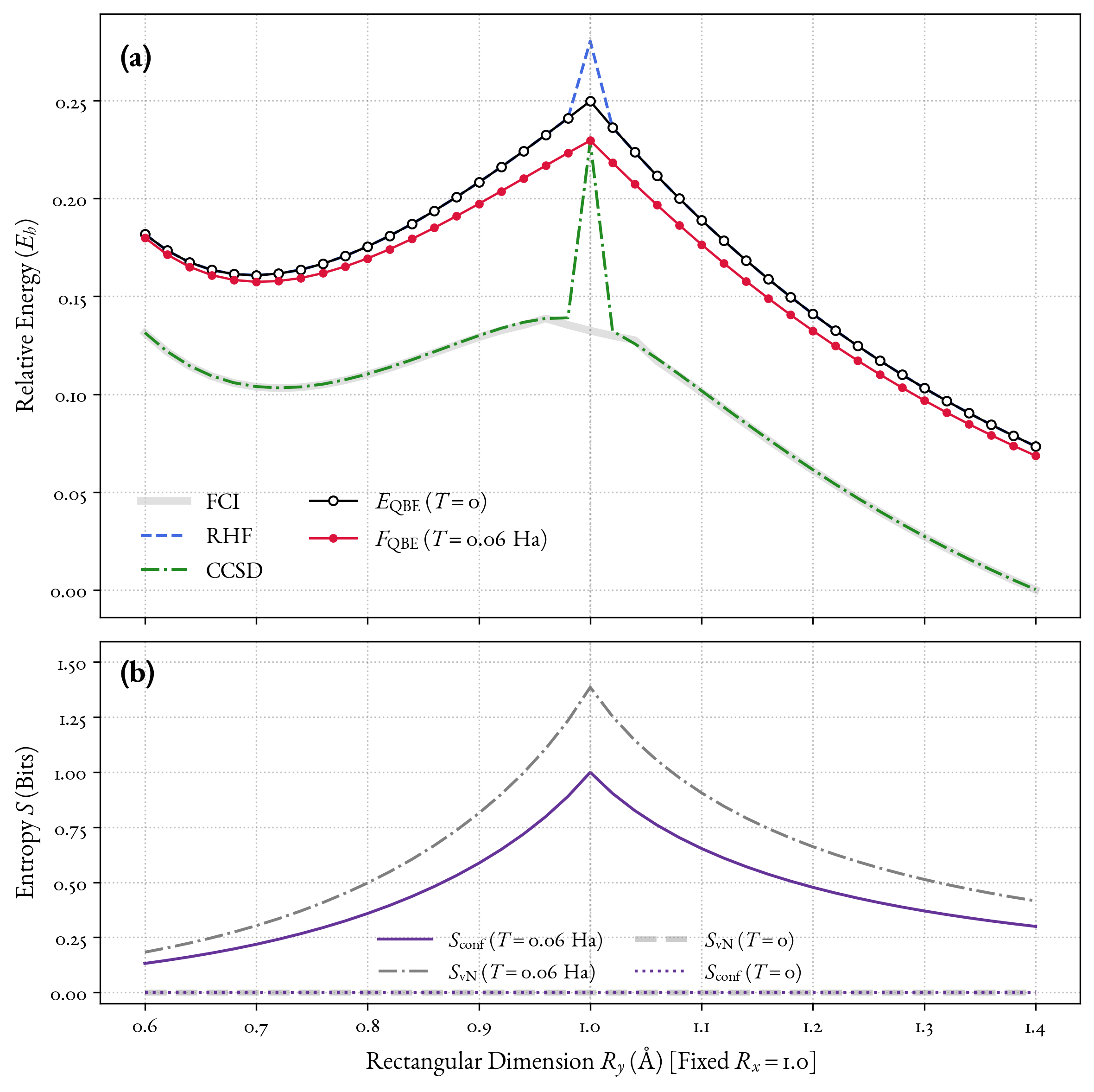}
        \caption{\textbf{H$_4$ D$_{\mathrm{2h}}\rightarrow$ D$_{4h}\rightarrow$ D$_\mathrm{4h}$ rectangular distortion in \texttt{6-31g}}. \textbf{(a)} QBE-SCF at $T=0$ bypasses the RHF discontinuity at $R_y = 1.0$ \AA. \textbf{(b)} The entropy is zero at $T=0$. A temperature of $T=0.06$ Ha induces fractionalization at the crossing, lowering the free energy.}
        \label{fig:h4_rectangular}
    \end{minipage}\hfill
    \begin{minipage}[t]{0.48\textwidth}
        \centering
        \includegraphics[width=\linewidth]{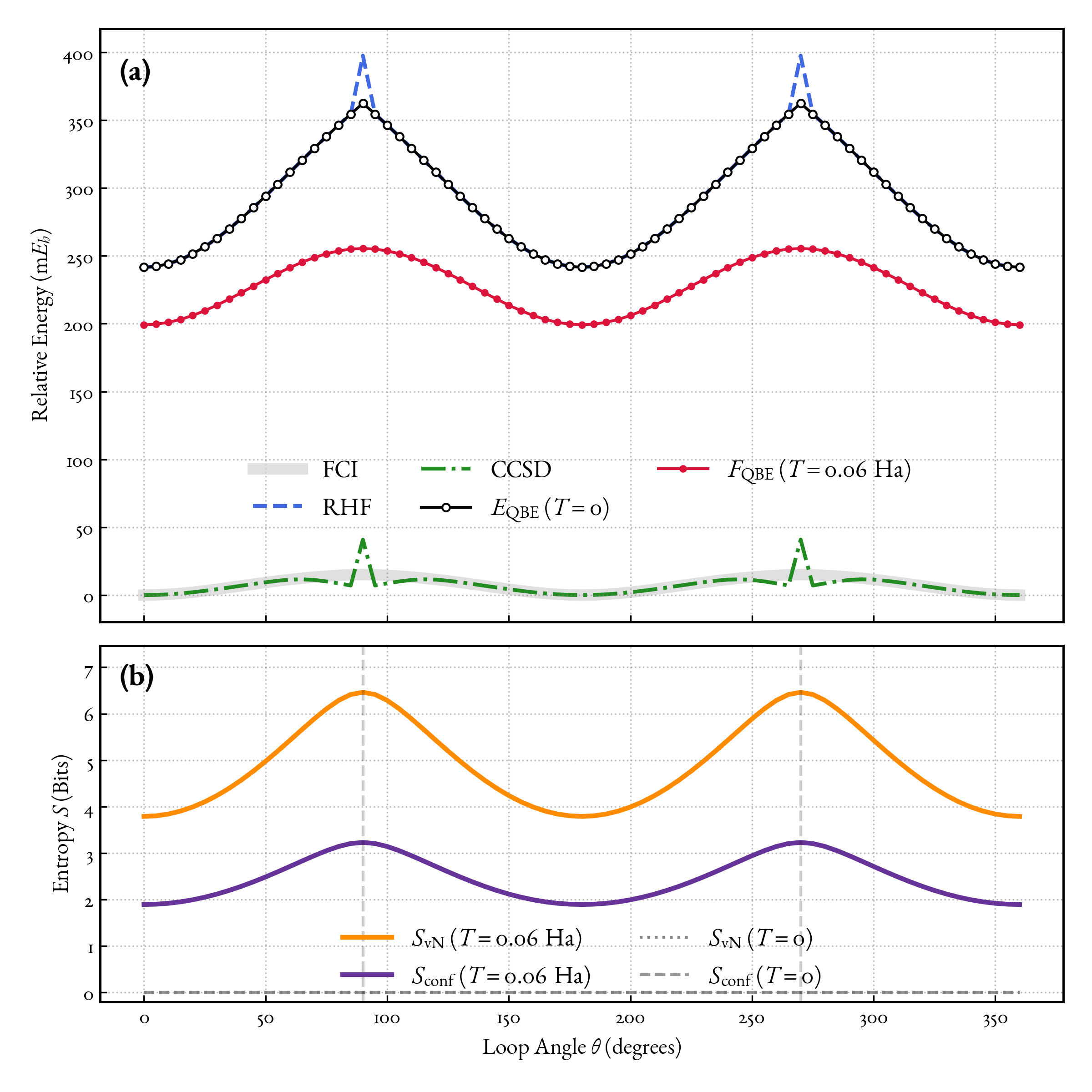}
        \caption{\textbf{H$_4$ Berry loop around the $D_{4h}$ conical intersection}. \textbf{(a)} $E_{QBE}(T=0)$ bypasses the RHF discontinuities at $\theta = 90^\circ$ and $270^\circ$. Applying an electronic temperature of $T=0.06$ Ha yields a differentiable free energy profile ($F_{\mathrm{QBE}}$) where thermal disorder ($S_{\mathrm{vN}(T)}$) regularizes the free energy.}
        \label{fig:h4_berry_loop}
    \end{minipage}
\end{figure*}

Figure \ref{fig:h4_rectangular} displays the coordinate stretch results. The RHF reference exhibits a cusp at $R_y = 1.0\ \text{\AA}$. The zero-temperature internal energy $E_{\mathrm{QBE}}(T=0)$ bypasses this cusp, maintaining zero entropy across the coordinate. Applying a finite electronic temperature of $T_{\mathrm{elec}} = 0.06\ E_h$ induces spatial fractionalization at the crossing. Minimizing the Helmholtz free energy utilizes this thermal disorder to stabilize the symmetric configuration, yielding a continuous crossover that parallels FCI.

To evaluate the structural consequence of this regularization, we examine the geometric phase along a closed angular contour, $\theta \in [0, 360^\circ]$, encircling the degeneracy at a constant radius of $r = 0.2\ a_0$. A real-valued state traversing a conical intersection accumulates a Berry phase of $\gamma = \oint \langle \Psi | \nabla \Psi \rangle \cdot d\mathbf{R} = -\pi$.~\cite{longuet1958studies, mead1979determination, berry1984quantal} In the phase-space formulation, the global scalar phase vanishes within the outer product of the density matrix: $\mathbf{P} = \left(e^{i\gamma}|\Psi\rangle\right)\left(e^{-i\gamma}\langle\Psi|\right) = |\Psi\rangle\langle\Psi|$. Thus, the topological invariant is preserved not by a discontinuous scalar phase flip, but through the continuous spatial rotation of the fractionally occupied active space.

Figure~\ref{fig:h4_berry_loop} depicts the relative energy and entropy profiles along this closed contour. The RHF and CCSD energies exhibit sharp derivative discontinuities at the $D_{4h}$ transition states ($\theta = 90^\circ$ and $270^\circ$), which are avoided by the $E_{\mathrm{QBE}}(T=0)$. Furthermore, applying an electronic temperature of $0.06\ E_h$ yields a macroscopic free energy profile, $F_{\mathrm{QBE}}$, that maintains a continuously differentiable surface congruent with the exact FCI adiabatic ground state. As demonstrated in Panel (b), this smoothing is thermodynamic in origin. The elevated thermal disorder ($S_{\mathrm{vN}}$) acts as a driving force, inducing continuous spatial fractionalization ($S_{\mathrm{conf}}$) of the frontier orbital occupations before the nuclei reach the exact geometric degeneracies. Taken together, these results indicate a duality in the solver's regularization mechanics: whereas the topological singularities in H$_2$ dissociation and symmetric $D_{3h}$ expansion of H$_3$ are repaired by kinematic ergodicity along the zero-temperature optimization pathway, the structural invariant of the H$_4$ conical intersection is preserved through thermodynamic regularization.

\begin{figure*}[t!]
    \centering
    \includegraphics[width=0.95\textwidth]{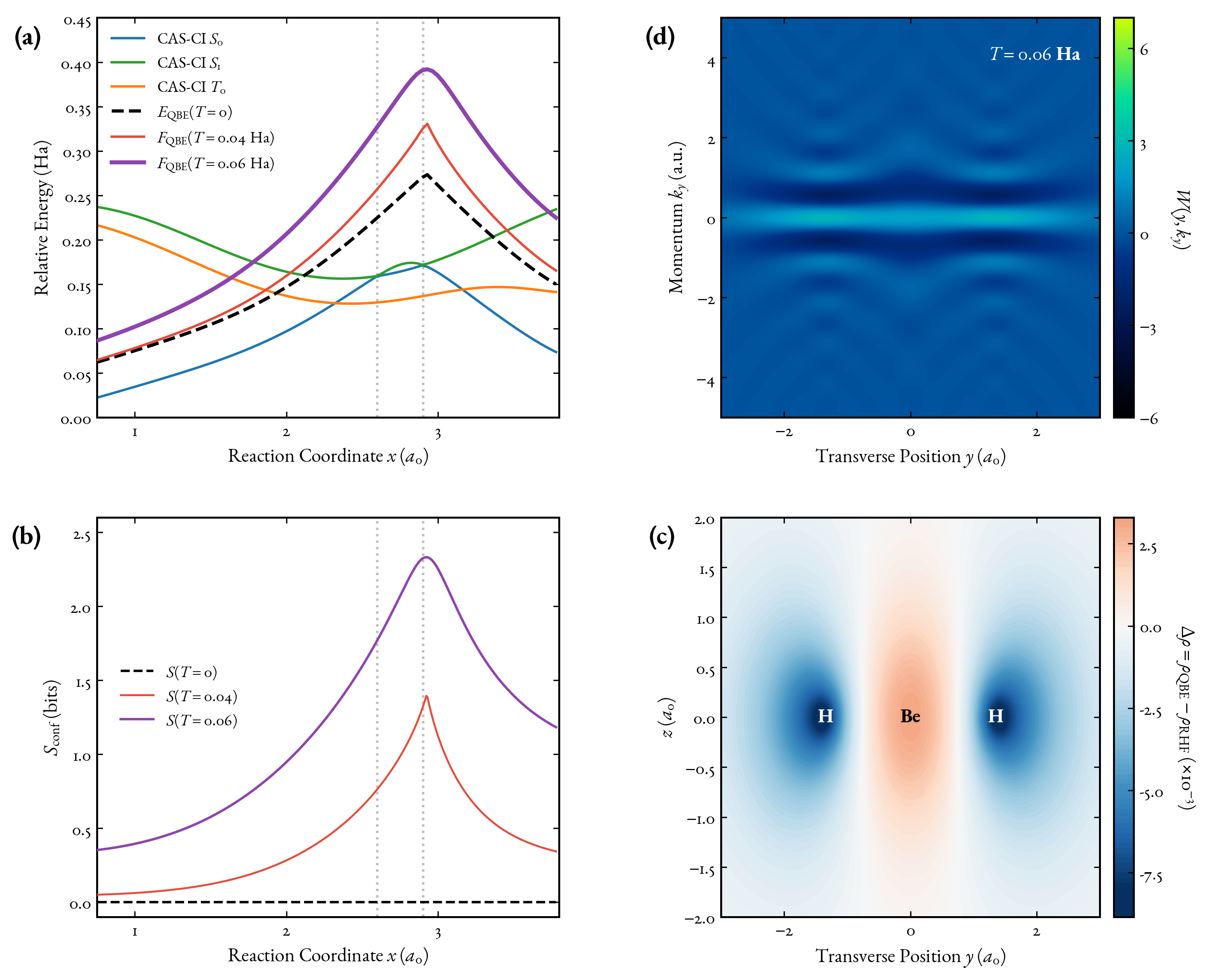}
    \caption{\textbf{Entropic regularization of the BeH$_2$ conical intersection in \texttt{6-31g}.} \textbf{(a)} Energies evaluated against the CAS-CI(4,12) linear limit. $E_{\mathrm{QBE}}(T=0)$ exhibits a sharp cusp and thermal entropy regularizes the avoided crossing. \textbf{(b)} Configurational entropy profiles confirm that the regularization of the singularity in (a) is driven by a peak in occupational fractionalization at the geometric degeneracy. \textbf{(c)} The spatial density difference ($\Delta \rho = \rho_{\mathrm{QBE}} - \rho_{\mathrm{RHF}}$) sliced along the transverse $y-z$ plane at $x = 2.6\ a_0$ confirms that QBE-SCF maintains spatial symmetry consistent with an RHF description. \textbf{(d)} Wigner distribution at the conical intersection ($x = 2.6\ a_0$, $T=0.06\ \mathrm{Ha}$) shows quantum fluctuations that drive the spatial stabilization.}
    \label{fig:beh2_phase_space}
\end{figure*}

\subsection{The \texorpdfstring{$\text{BeH}_2$}{BeH2} Conical Intersection}

We examine the $C_{2v}$ conical intersection of beryllium hydride ($\text{BeH}_2$) during the insertion of a Be atom into an $\text{H}_2$ molecule. We employ the parameterization of Purvis and Bartlett.~\cite{purvis1982full} The Be atom is fixed at the origin, with H atoms positioned at $(x, \pm y, 0)$ in Bohr ($a_0$), where the transverse internuclear coordinate is defined as $y = 2.540 - 0.46 x$. Figure \ref{fig:beh2_phase_space} details the entropic regularization of this intersection.

Panel \ref{fig:beh2_phase_space}(a) shows the relative energy profiles. CAS-CI(4,12) calculations establish the avoided crossing of the $S_0$ and $S_1$ states alongside the $T_0$ triplet. The QBE-SCF trace at $T=0$ exhibits a cusp at $x = 2.6\ a_0$. This arises because the single-determinant solution becomes trapped in a specific diabatic configuration, failing to mix orbital character across the intersection. While complex-valued Restricted Hartree-Fock (cRHF) can smooth this singularity by breaking complex conjugation symmetry,~\cite{Small2015} the resulting wavefunction remains an integer-occupied, zero-entropy pure state. In contrast, QBE-SCF restricts the atomic orbital basis to the real plane and resolves the intersection through entropic fractionalization. By smearing orbital occupations via Fermi-Dirac kinetics, our approach generates von Neumann entropy, recovering the multireference character of the exact state without complex arithmetic. To prevent the solver from trapping in the higher-energy diabatic state via density continuation, we filter forward and backward RHF sweeps to isolate the absolute energy minimum at each coordinate. At $T = 0.06\ \mathrm{Ha}$, this macroscopic free energy ($F_{\mathrm{QBE}}$) smoothly regularizes the discontinuity.

In panel \ref{fig:beh2_phase_space}(b) we plot the configurational entropy ($S_{\mathrm{conf}}$) along the reaction coordinate. The zero-temperature cusp is pinned to zero entropy. At finite temperatures, the macroscopic smoothing of the singularity is driven by a sharp peak in spatial fractionalization at the geometric degeneracy, thermodynamically offsetting the energy penalty of the mean-field cusp. Panel \ref{fig:beh2_phase_space}(c) maps the spatial density difference $\Delta \rho = \rho_{\mathrm{QBE}} - \rho_{\mathrm{RHF}}$ in the transverse $y-z$ plane, evaluated at the conical intersection geometry ($x = 2.6\ a_0$) for $T = 0.06\ \mathrm{Ha}$. The uniform charge density distribution confirms that QBE-SCF maintains spatial symmetry consistent with a restricted (RHF) description.

The Wigner distribution along the internuclear $y$-axis is shown in panel \ref{fig:beh2_phase_space}(d), also at the intersection geometry ($x = 2.6\ a_0$) for $T = 0.06\ \mathrm{Ha}$. A continuous central band at $k_y \approx 0$ confirms the restoration of a symmetric bonding network. The interference fringes at elevated momenta exhibit negative quasi-probability. These negative regions are the signature of non-classical electronic correlation. QBE-SCF recovers the correlation of the CAS-CI benchmark through finite-temperature phase-space transport. This entropic regularization complements the zero-temperature kinetic ergodicity that resolves the spatial degeneracy of $\text{H}_3$. Because the $\text{BeH}_2$ $C_{2v}$ insertion features an avoided crossing of states with identical symmetry ($^1A_1$) rather than a strictly spatial degeneracy, thermal smearing is required to bypass the mean-field cusp and smoothly recover the correlated adiabatic surface.

\section{Conclusion}
We formulated a Quantum Boltzmann Equation (QBE) for molecular electronic structure in which the projected Wigner distribution is represented by the one-particle density matrix ${}^{1}\mathbf{P}$ in an atomic-orbital basis and evolved by BGK relaxation toward a Fermi-Dirac distribution defined by the instantaneous Fock matrix. At stationarity QBE-SCF enforces $[\mathbf{F},\mathbf{P}] = \mathbf{0}$, i.e., the Hartree-Fock self-consistency condition in the $T_{elec} \to 0$ limit which reduces to the idempotent Aufbau projector for H$_{2}$ at equilibrium. When initialized with a perturbed superposition of atomic densities (SAD) broadcast to an unrestricted density matrix, this zero-temperature advection navigates the Coulson-Fischer instability to recover the UHF minimum in H$_2$. Similarly, seeding the solver with a spin-perturbed guess captures the generalized valence bond (GVB) limit during H$_3$ dissociation in D$_{3h}$.

A practical consequence of the kinetic formalism is its capacity to resolve mean-field singularities through zero-temperature kinetic ergodicity or finite-temperature thermodynamic regularization. For systems exhibiting true spatial degeneracies, such as the symmetric dissociation in H$_3$ in D$_\mathrm{3h}$, the Markovian collision operator fractionalizes the density matrix at $T_{\mathrm{elec}}=0$. By kinetically averaging over broken-symmetry states, the solver bypasses single-determinant derivative discontinuities to converge to the Full CI limit.

Conversely, for the $\text{BeH}_{2}$ insertion coordinate, the framework resolves the mean-field cusp via finite-temperature free-energy optimization, $F = E - T_{\mathrm{elec}}S_{\mathrm{vN}}$. Because this $C_{2v}$ path features an avoided crossing of identically symmetric states rather than a true topological degeneracy, thermal smearing is required to bridge the diabatic states. Unlike complex-valued Restricted Hartree-Fock, which smooths this cusp by breaking complex conjugation symmetry to yield an integer-occupied pure state, QBE-SCF utilizes fractional occupations. The resulting von Neumann entropy acts as a thermodynamic order parameter that stabilizes the symmetric mixed state, recovering multi-reference character within a real-valued single-determinant basis. 

Extending this entropic regularization to the $D_{4h}$ conical intersection of H$_4$ illustrates how continuous fractionalization preserves topological invariants. When traversing a closed angular loop around the degeneracy, integer-constrained methods exhibit derivative discontinuities due to abrupt spatial orbital permutations. At finite temperatures, QBE-SCF resolves these singularities, yielding a differentiable free energy profile congruent with the adiabatic ground state. Thermal disorder drives a continuous spatial fractionalization of the frontier orbitals, allowing the geometric Berry phase of the intersection to be preserved through the spatial rotation of the active space rather than a discontinuous scalar phase flip.

These results demonstrate that formulating self-consistent-field optimization as a kinetic relaxation offers the synthesis of single-reference electronic structure theory and quantum statistical mechanics. By utilizing Markovian density matrix damping  QBE-SCF captures signatures of strong static correlation within a mean-field paradigm. Important next steps are to (i) quantify how the optimal $T_{elec}$ scales with gap size and basis completeness, (ii) extend the collision model beyond BGK to incorporate correlation effects beyond Hartree-Fock, and (iii) couple the resulting regularized electronic surfaces to nuclear dynamics and nonadiabatic observables, offering a phase-space alternative to nuclear velocity rescalings required by traditional energy-conserving trajectory surface hopping methods.~\cite{tully1990molecular}

\section*{SUPPLEMENTARY MATERIAL}

See the supplementary material for additional theoretical analysis,
numerical details, convergence data, and supporting results.

% ============================================================
% END MATTER
% ============================================================

\section*{Acknowledgments}

The author thanks Professor Martin Head-Gordon of the University of California, Berkeley, for helpful discussions during the early development of the ideas underlying this work.

\section*{Author Declarations}

\subsection*{Conflict of Interest}

Romit Chakraborty is the founder and an equity holder of Point Reyes Sound, Inc., which serves as the assignee of U.S. Provisional Patent Application No.~64/033,274 related to this work.

\subsection*{Author Contributions}

\textbf{Romit Chakraborty:} Conceptualization, Methodology, Software, Formal analysis, Investigation, Visualization, Writing--original draft, Writing--review and editing.

\vspace{1em}

\section*{Data Availability}
The data that support the findings of this study are openly available in Zenodo at \url{https://doi.org/10.5281/zenodo.21940451}. 
% Algorithmic implementation details are available from the corresponding author upon reasonable request.

\section*{References}

\bibliographystyle{aipnum4-2}

%aipnum4-2.bst 2019-01-14 (MD) hand-edited version of apsrev4-1.bst

\clearpage
\onecolumngrid  % <--- Forces full-width single-column layout for the SI

% --- SI Formatting & Counter Resets ---
\renewcommand{\thesection}{S\Roman{section}}
\renewcommand{\thesubsection}{\Alph{subsection}}
\renewcommand{\thefigure}{S\arabic{figure}}
\renewcommand{\thetable}{S\arabic{table}}
\renewcommand{\theequation}{S\arabic{equation}}

\setcounter{section}{0}
\setcounter{figure}{0}
\setcounter{table}{0}
\setcounter{equation}{0}

% --- SI Title Block ---
\begin{center}
{\large\bfseries Supplementary Material for\par}
\vspace{0.8em}
{\LARGE\bfseries Quantum Boltzmann Equation Self-Consistent-Field for the Entropic Regularization of Mean-Field Singularities\par}
\vspace{1.2em}
{\large Romit Chakraborty\par}
\vspace{0.35em}
{\itshape Point Reyes Sound, Inc., San Francisco, California 94114, USA\par}
\vspace{0.35em}
{\small Correspondence: \href{mailto:romit@pointreyessound.com}{romit@pointreyessound.com}\par}
\end{center}

\vspace{1em}
\hrule
\vspace{1.5em}
\section{DISCRETIZATION OF THE QUANTUM BOLTZMANN EQUATION}

We detail here the phase-space projection and discretization of the continuous Quantum Boltzmann Equation (QBE) into a discrete matrix propagation scheme. 

\subsection*{A. Phase-Space Projection and the Wigner Transform}
The fundamental object of our phase-space mechanics is the Wigner quasi-probability distribution, $W(\mathbf{r}, \mathbf{p}, t)$.~\cite{SI_wigner1932, SI_hillery1984} To render this continuous function computationally tractable, we project $W(\mathbf{r},\mathbf{p},t)$ onto a finite basis of atomic orbitals, $\{\chi_\mu\}$. We first define the Wigner transform of the static orbital dyad, $\chi_\mu(\mathbf{r})\chi_\nu^*(\mathbf{r}')$. Working in atomic units ($\hbar = 1, m_e = 1, e = 1$), this transform evaluates the non-local spatial overlap over a relative coordinate $\mathbf{s}$:
\begin{equation}
\mathcal{W}_{\mu\nu}(\mathbf{r}, \mathbf{p}) = \frac{1}{(2\pi)^3} \int e^{-i \mathbf{p} \cdot \mathbf{s}} \chi_\mu\left(\mathbf{r} + \frac{\mathbf{s}}{2}\right) \chi_\nu^*\left(\mathbf{r} - \frac{\mathbf{s}}{2}\right) d^3s.
\end{equation}
Because the underlying Gaussian basis functions are locked to stationary nuclear centers, the resulting phase-space basis functions $\mathcal{W}_{\mu\nu}(\mathbf{r}, \mathbf{p})$ are static in time. We construct the macroscopic fluid by tracing these static basis functions against the time-dependent one-particle density matrix, $P_{\mu\nu}(t)$:
\begin{equation}
W(\mathbf{r},\mathbf{p},t) = \sum_{\mu\nu} P_{\mu\nu}(t) \mathcal{W}_{\mu\nu}(\mathbf{r},\mathbf{p}).
\end{equation}
Because the basis is static, any temporal evolution of the phase-space fluid is entirely isolated within the density matrix expansion coefficients, $\frac{\partial P_{\mu \nu}}{\partial t}$.

\subsection*{B. Continuous Advection and the Moyal Star Product}
The evolution of the fluid is governed by the integro-differential QBE:
\begin{equation}
\frac{\partial W}{\partial t} + \frac{\mathbf{p}}{m} \cdot \nabla_\mathbf{r} W + \mathcal{F} \cdot \nabla_\mathbf{p} W = \mathcal{Q}[W]
\end{equation}
In classical mechanics, spatial streaming ($\frac{\mathbf{p}}{m} \cdot \nabla_\mathbf{r} W$) and momentum drift ($\mathcal{F} \cdot \nabla_\mathbf{p} W$) constitute the Poisson bracket, $-\{H, W\}_{\text{PB}}$. In quantum mechanics, this spatial and momentum advection is generalized to the Moyal bracket, $\{H, W\}_{\text{MB}}$.~\cite{SI_moyal1949} The Moyal bracket is formally defined via the phase-space Moyal star product ($\star$), which encapsulates the infinite-order gradient expansions of quantum dispersion. In atomic units, this evaluates to $\{H, W\}_{\text{MB}} = -i(H \star W - W \star H)$.~\cite{SI_zachos2005} 

The fundamental theorem of the Wigner-Weyl transform establishes that the star product of functions in phase space maps identically to the standard multiplication of operators in Hilbert space.\cite{SI_wigner1932, SI_hillery1984} Consequently, the continuous Moyal bracket maps exactly to the discrete matrix commutator. Applying the mean-field approximation (replacing the exact many-body Hamiltonian with the effective single-particle Fock operator, $F = T + V_{\text{eff}}$), the continuous advection maps to the discrete matrix commutator:
\begin{equation}
\frac{\partial \mathbf{P}}{\partial t} - i[\mathbf{F}, \mathbf{P}] = \mathcal{Q}[\mathbf{P}]
\end{equation}
The linearity of the commutator explicitly partitions the advection: the spatial streaming maps to the kinetic energy commutator ($-i[\mathbf{T}, \mathbf{P}]$) driving quantum dispersion, and the momentum drift maps to the potential energy commutator ($-i[\mathbf{V}_{\text{eff}}, \mathbf{P}]$) driving dynamic acceleration.

\subsection*{C. Phenomenological Discretization}
Employing the Bhatnagar-Gross-Krook (BGK) relaxation time approximation for the collision operator $\mathcal{Q}[\mathbf{P}]$, we model collisions as a continuous relaxation toward a local target equilibrium, $\mathbf{P}^{\mathrm{eq}}$, over a relaxation time $\tau$:~~\cite{SI_bgk1954}
\begin{equation}
\frac{\partial \mathbf{P}}{\partial t} - i[\mathbf{F}, \mathbf{P}] = -\frac{1}{\tau} (\mathbf{P} - \mathbf{P}^{\mathrm{eq}})
\end{equation}
To isolate the stationary ground state (where $[\mathbf{F}, \mathbf{P}] = \mathbf{0}$) and avoid non-dissipative Rabi oscillations, we drop the conservative commutator $-i[\mathbf{F}, \mathbf{P}]$ during the micro-step update. Discretizing the resulting differential equation using a forward Euler finite difference over a time step $\Delta t$ yields:
\begin{equation}
\frac{\mathbf{P}(t+\Delta t) - \mathbf{P}(t)}{\Delta t} = -\frac{1}{\tau} (\mathbf{P}(t) - \mathbf{P}^{\mathrm{eq}}[\mathbf{F}(t)])
\end{equation}
Rearranging these terms and defining the dimensionless collision frequency as $\omega = \frac{\Delta t}{\tau}$, we arrive directly at the QBE-SCF matrix propagation scheme:
\begin{equation}
\mathbf{P}(t+\Delta t) = (1-\omega)\mathbf{P}(t) + \omega \mathbf{P}^{\mathrm{eq}}[\mathbf{F}(t)]
\label{eq:bgk_si}
\end{equation}

\section{Theoretical Equivalence to the Variational Limit}

In this section, we demonstrate that the stationary state of the Boltzmann Equation in the basis of atomic orbitals (AOs) is equivalent to the self-consistent solution of the Roothaan-Hall equations (Restricted Hartree-Fock limit). 

The evolution of the electronic density matrix $\mathbf{P}(t)$ is governed by the discrete-time BGK collision operator derived in Equation \ref{eq:bgk_si}, where $\omega \in (0, 2)$ is the relaxation frequency and $\mathbf{F}(t) = \mathbf{H}_{\mathrm{core}} + \mathbf{J}[\mathbf{P}(t)] - \mathbf{K}[\mathbf{P}(t)]$ is the instantaneous Fock matrix.

\subsection{The Stationarity Condition}
The solver reaches a steady state when the density matrix ceases to evolve, i.e., $\mathbf{P}(t+\Delta t) = \mathbf{P}(t) = \mathbf{P}_{stat}$. Substituting this into Eq. (\ref{eq:bgk_si}) yields:
\begin{equation}
    \mathbf{P}_{stat} = (1-\omega)\mathbf{P}_{stat} + \omega \mathbf{P}^{eq}[\mathbf{F}(\mathbf{P}_{stat})] \implies \mathbf{P}_{stat} = \mathbf{P}^{eq}[\mathbf{F}_{stat}]
\end{equation}
Thus, at any converged fixed point, the density matrix equals the
equilibrium distribution defined by its self-consistent Fock matrix.

\subsection{Commutation and Variational Optimality}
The equilibrium distribution $\mathbf{P}^{eq}$ is constructed via the Fermi-Dirac statistics of the Fock matrix:
\begin{equation}
    \mathbf{P}^{eq} = f_{FD}(\mathbf{F}, \mu, \beta) = \left[ \mathbf{I} + \exp(\beta(\mathbf{F} - \mu\mathbf{I})) \right]^{-1}
\end{equation}
Since any matrix function $f(\mathbf{A})$ commutes with its argument $\mathbf{A}$, it follows that:
\begin{equation}
    [\mathbf{F}_{stat}, \mathbf{P}_{stat}] = 0
\end{equation}
This commutation relation is the exact Hartree-Fock stationarity condition. Unlike conventional repeated-diagonalization schemes where stationary states may correspond to saddle points, the dissipative nature of the BGK collision operator continuously relaxes the fluid along the gradient of the free energy. Provided the initial state contains sufficient spatial noise to break artificial symmetries, this thermodynamic relaxation autonomously drives the system away from unstable stationary points and toward a stable energetic minimum.

\section{Validation of the Coulson-Fischer Point}

To validate the phase-space dynamics of the Quantum Boltzmann Equation self-consistent-field (QBE-SCF) solver, we compared the location of the Coulson-Fischer (CF) point ($R_{\mathrm{CF}}$) derived from our framework against the exact energetic bifurcation point obtained from static Hartree-Fock theory. The CF point represents a continuous phase transition where the restricted electronic Hessian develops a negative eigenvalue. Near this bifurcation, the potential energy surface becomes asymptotically flat ($\nabla E \to 0$). Because QBE-SCF utilizes physical Markovian density matrix damping rather than mathematical extrapolation (e.g., DIIS), the kinetic fluid natively experiences \textit{critical slowing down}. To rigorously evaluate this, we define the transition through two distinct limits:

\begin{enumerate}
    \item \textbf{Kinetic Ergodicity ($R_{\mathrm{Kin}}$):} Evaluated strictly at zero electronic temperature ($T_{\mathrm{elec}} = 0$). The solver is initialized from a raw, trace-conserving spin-perturbed Superposition of Atomic Densities (SAD) guess. The critical point is identified by the maximum spatial gradient of the configurational entropy ($S_{\mathrm{config}}$), representing the exact moment the zero-temperature advection navigates the flat potential energy surface to spontaneously break symmetry. To allow the physical fluid to resolve the topological stiffness near the bifurcation, the maximum solver cycle limit was extended to 5000.
    \item \textbf{Thermodynamic Probe ($R_{\mathrm{Thm}}$):} Evaluated using a vanishingly small electronic temperature ($T_{\mathrm{elec}} = 0.002$ Ha) acting strictly as a gap probe. The solver is seeded with the fully converged unrestricted Hartree-Fock (UHF) density matrix. The transition is identified by the gradient of the von Neumann thermal entropy ($S_{\mathrm{vN}}$), which sharply registers the exact geometric threshold where the single-particle HOMO-LUMO gap closes.
\end{enumerate}

\noindent Table~\ref{tab:cf_validation_unified} presents the results evaluated on a high-density spatial grid ($\Delta R = 0.001$ \AA). The phase-space framework exhibits exceptional agreement with the energetic ground truth. For the correlation-consistent basis sets (\texttt{cc-pVDZ}, \texttt{cc-pVTZ}, and \texttt{aug-cc-pVTZ}), the zero-temperature kinetic advection perfectly identifies the exact geometric bifurcation point ($|\Delta_{\mathrm{Kin}}| = 0.0000$ \AA). Furthermore, the cycle counts record the physical signature of the continuous phase transition. While the thermodynamic probe rapidly confirms the converged electronic state, the kinetic solver requires thousands of steps exactly at the divergence threshold, demonstrating the critical slowing down of the phase-space fluid as the restoring forces vanish.

\begin{table}[h!]
    \centering
    \caption{\textbf{Thermodynamic and kinetic validation of the Coulson-Fischer point.} Comparison of the critical bond length ($R_{\mathrm{CF}}$) for H$_2$ computed via exact static energetics ($R_{\mathrm{True}}$), zero-temperature kinetic ergodicity ($R_{\mathrm{Kin}}$), and the finite-temperature thermodynamic probe ($R_{\mathrm{Thm}}$). $\Delta$ denotes the absolute error relative to $R_{\mathrm{True}}$. The step counts represent the SCF cycles required at the bifurcation point ($\Delta R = 0.001$ \AA\ grid resolution).}
    \label{tab:cf_validation_unified}
    \vspace{0.2cm}
    \renewcommand{\arraystretch}{1.25} % Gives the rows elegant vertical breathing room
    \begin{tabular*}{\textwidth}{@{\extracolsep{\fill}} l c c c c c c c @{}}
        \toprule
        & & \multicolumn{3}{c}{\textbf{$T = 0$}} & \multicolumn{3}{c}{\textbf{$T = 0.002$} Ha} \\
        \cmidrule(lr){1-2} \cmidrule(lr){3-5} \cmidrule(lr){6-8}
        \textbf{Basis Set} & \textbf{$R_{\mathrm{True}}$ (\AA)} & \textbf{$R_{\mathrm{Kin}}$ (\AA)} & \textbf{$|\Delta_{\mathrm{Kin}}|$ (\AA)} & \textbf{Cycles} & \textbf{$R_{\mathrm{Thm}}$ (\AA)} & \textbf{$|\Delta_{\mathrm{Thm}}|$ (\AA)} & \textbf{Cycles} \\
        \midrule
        STO-3G      & 1.1540 & 1.1560 & 0.0020 & 4505 & 1.2130 & 0.0590 & 494 \\
        6-31G       & 1.1920 & 1.1930 & 0.0010 & 5000 & 1.1930 & 0.0010 & 2544 \\
        cc-pVDZ     & 1.2120 & 1.2120 & 0.0000 & 5000 & 1.2120 & 0.0000 & 1805 \\
        cc-pVTZ     & 1.2160 & 1.2160 & 0.0000 & 5000 & 1.2160 & 0.0000 & 1 \\
        aug-cc-pVTZ & 1.2180 & 1.2180 & 0.0000 & 5000 & 1.2180 & 0.0000 & 1 \\
        \bottomrule
    \end{tabular*}
\end{table}

\section{Entropic Regularization At Conical Intersections}
For systems exhibiting near-degeneracy, such as the $2a_1$ and $1b_2$ orbitals in the transition state of $\text{BeH}_2$, the zero-temperature limit ($\beta \to \infty$) leads to a convergence failure or unphysical energy cusp in single-reference theories. We resolve this singularity by treating the electronic fluid as a canonical ensemble with finite inverse temperature $\beta < \infty$. The stationarity condition $\mathbf{P}_{stat} = f_{\mathrm{FD}}(\mathbf{F}_{stat}, \beta)$ remains valid, but the resulting density matrix becomes:
\begin{equation}
    \mathbf{P}_{stat} = \sum_{i} n_i(\epsilon_i, \beta) | \psi_i \rangle \langle \psi_i |
\end{equation}
where $n_i = \left[1 + \exp(\beta(\epsilon_i - \mu))\right]^{-1}$ are fractional occupation numbers. The chemical potential $\mu$ is determined iteratively at each step to enforce particle conservation, $\text{Tr}(\mathbf{P}\mathbf{S}) = N_{\text{elec}}$. This framework effectively corresponds to Finite-Temperature Hartree-Fock (FTHF) theory. By allowing the electron fluid to populate the degenerate manifold entropically, the solver minimizes the Helmholtz Free Energy ($F = E - T_{\mathrm{elec}}S_{\mathrm{vN}}$) rather than the internal energy $E$, transforming the non-differentiable energy cusp into a smooth thermodynamic crossover.

\vspace{2em}
\hrule

\end{document}